\documentclass[12pt, a4paper, oneside]{article}

\usepackage[a4paper, tmargin=1.1in, lmargin=1in, rmargin=1in, bmargin=1.1in]{geometry}

\usepackage{setspace}
\usepackage{graphicx}
\usepackage{booktabs,array,multirow}
\usepackage{url}
\usepackage{hyperref}
\usepackage[utf8]{inputenc}
\usepackage[english]{babel}
\usepackage{siunitx}
\usepackage{bbm}
\usepackage{amsfonts}
\usepackage{amsmath,amssymb}
\usepackage{amsthm}
\usepackage{comment}
\usepackage{makecell}
\usepackage{rotating}
\usepackage{dsfont} 
\usepackage{float} 
\usepackage{setspace}
\usepackage{natbib}

\newif\iflatexml\latexmlfalse

\AtBeginDocument{\DeclareGraphicsExtensions{.pdf,.PDF,.eps,.EPS,.png,.PNG,.tif,.TIF,.jpg,.JPG,.jpeg,.JPEG}}

\DeclareMathOperator{\EX}{\mathbb{E}}% expected value

\newtheorem{proposition}{Proposition}

\title{Mind the Income Gap: Bias Correction of Inequality Estimators in Small-Sized Samples}

\author{Silvia De Nicolò\thanks{\texttt{silvia.denicolo@unibo.it}} \and Maria Rosaria Ferrante \and Silvia Pacei}

\date{%Department of Statistical Sciences ``P. Fortunati",
University of Bologna}

\begin{document}

\newcommand\barbelow[1]{\stackunder[1.2pt]{$#1$}{\rule{.8ex}{.075ex}}}

\maketitle
\selectlanguage{english}
\begin{abstract}
\noindent Income inequality estimators are biased in small samples, leading generally to an underestimation. This aspect deserves particular attention when estimating inequality in small domains and performing small area estimation at the area level. % After investigating the nature of the bias,
We propose a bias correction framework for a large class of inequality measures comprising the Gini Index, the Generalized Entropy and the Atkinson index families by accounting for complex survey designs. The proposed methodology does not require any parametric assumption on income distribution, being very flexible. Design-based performance evaluation of our proposal has been carried out using EU-SILC data, their results show a noticeable bias reduction for all the measures. Lastly, an illustrative example of application in small area estimation confirms that ignoring ex-ante bias correction determines model misspecification.
\end{abstract}%



\section{Introduction}

The interest in reliable local estimates of economic inequality is growing due to the observed increment in the income gap and social exclusion among regions. Specifically, inequality estimates for specific sub-populations - such as areas at a fine level of geographical disaggregation or rather specific socio-demographic groups - are increasingly in demand \citep{marquez2019role}. Policymakers and stakeholders need these to formulate and implement policies, distribute resources and measure the effect of policy actions at local levels. In addition, their contribution to regional studies is valuable in the process of decomposing spatial spillovers and identifying local areas that drive inequality at national levels \citep{cavanaugh2018locating}.

When dealing with inequality estimation in specific groups or local scales, a problem of observations scarcity typically arises. Disposable income is generally adopted as the variable of interest and the primary source of data collection is through household
surveys. However, since such surveys are not planned
for the estimation of target quantities in specific domains, they result in small sample sizes. %Moreover, they are often subjected to data disclosure issues. 
 In this context, small area estimation techniques are applied, integrating survey with auxiliary data to "borrow strength" across areas and, in this way, improve the reliability of estimates. 
 
The small area models can be specified at the unit (individual or household) level; previous proposals dealing with inequality estimation at the unit level are provided by \citet{tzavidis2016robust} and \citet{marchetti2021robust}. However, such models require a large amount of data as, generally, the auxiliary variables have to be known for each unit of the population and linked to survey data. This may be hard to get as administrative archives are not publicly accessible at individual level, cross-linked and associated with survey data \citep{harmeningframework}. 
On the other hand, small area models defined at the area level are less demanding in terms of data requirements, needing only survey (direct) estimates endowed with related measures of uncertainty and areal covariates \citep[ch.4]{rao2015}. An application of such models to inequality estimation can be found in  \citet{benedetti2023}.
 
Area-level models with common specifications, such as the famous Fay-Herriot model \citep{fay1979estimates}, have the strict assumption of (approximate) unbiasedness of the survey estimators given as input \citep[ch.4]{rao2015}. 
In this paper, we focus on tha bias affecting inequality estimators in small samples, often underestimating inequality \citep{Deltas2003, Breunig2008}. Their bias may depend on the characteristic of the distribution of the variable of interest, i.e. the income variable \citep{Breunig2001}, and on the uncertainty induced by the sample selection scheme.  %for some specific measures, also on the skewness of its distribution \citep{Breunig2001}. %In this regard, consider that income is well-known to be positively skewed.
%Moreover, its magnitude depends on the type of measure considered. 

Unfortunately, the bias issue is typically neglected when measuring inequality with area-level models, leading to model misspecification and thus to a possible misleading inference. 
Note that such an aspect deserves attention given that estimates of inequality measures are often used for comparisons across time and locations. Neglecting such bias may bring out discrepancies that, rather than being true inequality gaps, may be due to disparate sample sizes or to different underlying distributions of the variable of interest \citep{Breunig2008}. In this vein, we propose a bias correction strategy for a large set of inequality measures and we adopt it in an illustrative small area estimation exercise. 

Concerning the Gini index, a large body of literature faces the small sample bias issue, such as \citet{jasso1979gini}, \citet{lerman1989improving},  \citet{Deltas2003}, \citet{Davidson2009}, \citet{Ourti2011} in $iid$ samples. The context of application is varied, spanning from economic inequality to crime or concentration of scholarly citations \citep{mohler2019reducing, kim2020influence}. 
\citet{Fabrizi2016} tackle such an issue in the complex survey case and their correction is indeed considered within a small area estimation framework. However, concerning alternative measures such as Atkinson Indexes and the Generalized Entropy (GE) measures, the literature on bias is very scarce, even in the $iid$ case: some contributions are provided by \citet{Giles2005}, \citet{Schluter2009} and \citet{Breunig2008} by adopting different methodological approaches of correction.

Note that income data are collected through household surveys with complex sampling designs that adopts stratification and/or selection of sampling units in more than one stage. Thus, the sample selection process, together with ex-post treatment procedures such as calibration and imputation, invariably introduces a complex correlation structure in the data that has to be taken into account. This makes the development of a theoretically valid bias correction challenging, in contrast to classical $iid$ settings. Furthermore, the bias issue is even exacerbated in income data applications, traditionally affected by extreme values \citep{Kerm2007}, since inequality measures are known to be highly unrobust to them \citep{cowell1996robustness}. 
This aspect depends clearly on the type of measure we are dealing with and it becomes even more cumbersome to handle in the case of small samples.

We investigate the nature of the bias and propose a methodological framework for bias correction. Our proposal constitutes a generalization of the framework of \citet{Breunig2008}, developed for $iid$ observations, to the finite population and design-based setting.  At the same time, we extend the proposal to a wider set of measures from the Gini index to two parametric families of measures: the Atkinson and the Generalized Entropy family. We considered a wide variety of measures as the concurrent estimation of alternative indicators - as opposed to the more commonly used Gini Index - may bring to light a wider picture of the inequality phenomenon. To the best of our knowledge, this is the first proposal of bias correction for the Atkinson and Generalized Entropy indexes in the complex survey case, whereas it provides an extension for the Gini index case with respect to existing proposals as it is made clear in Section \ref{biasest}.

To our purpose, we take advantage of a methodology based on Taylor's expansions, even if the same analytical results can be obtained through other types of linearization, such as the one proposed by \citet{graf2011use}. The extension for complex designs has been developed considering Horvitz-Thompson type estimators, and the ultimate clusters technique for design variances and covariances estimation.
An advantage of our proposal is that any parametric assumption on income distribution is not required, providing a very flexible framework. Our bias correction proposal is evaluated via simulations showing a noticeable bias reduction for all the measures and leading, in some cases, to approximately unbiased estimators. % Our results confirm the great impact extreme values have on the magnitude of estimator bias and error. 
Lastly, we provide a small area estimation exercise that shows the risk of ignoring ex-ante bias correction.

The paper is organized as follows. The considered inequality measures are defined in Section \ref{ineqmeasures}, while the bias correction strategy is set out in Section \ref{biasresults} and the bias-correction estimation steps are detailed in Section \ref{biasest}. A design-based simulation study involving the European Statistics on Income and Living Condition (EU-SILC) income data is provided in Section \ref{simulationsection} to evaluate the magnitude of the bias and the efficacy of our proposal. Lastly, a small area estimation exercise is carried out in Section \ref{smallareaexercise}, to highlight the utility of our proposal in practice.  Conclusions are drawn in Section \ref{conclusions}.

\section{Inequality Measures}
\label{ineqmeasures}
 
The most famous inequality measure is, indeed, the Gini concentration index,
employed in social sciences for measuring concentration in the distribution of a positive random variable. There are several equivalent definitions of the Gini index \citep{ceriani2015individual}, we use the formulation of \citet{sen1997economic}. Suppose we have a finite population $\mathcal{U}$ of $N (< \infty)$ elements labelled as $\lbrace 1, \dots, N \rbrace$. Let $y_i$ be a characteristic of interest, in our case income, for the $i$-th unit of the finite population,  where $y_i \in \mathbb{R}^+$, $i = 1, \dots, N$, and a sample $s_{iid}$ of size $n_{iid}$ is picked through simple random sampling. The Gini index estimator is defined as
\begin{align*}
\theta_G=\frac{2}{\hat{\mu} n_{iid}^2} \sum_{i \in s_{iid}} n_i y_i -\frac{n_{iid}+1}{n_{iid}},
\end{align*}
with $n_i$ denoting the rank of $i$-th unit and $\hat{\mu}$ the sample mean. %The Gini coefficient varies between 0 (case of perfect equality) and 1 (perfect inequality), and it is invariant under scale transformations. 

However, the estimation of alternative measures, in addition to the Gini index, may enable a more meaningful assessment of different aspects of economic inequality. The Gini index is decomposable within and between groups only in very specific cases \citep{mookherjee1982decomposition}, moreover, it is positional (weakly) transfer sensitive, namely because index variations induced by income transfers depend on the ranks of transfer donor and recipient. Lastly, it constitutes a Lorenz dominance-based measure, allowing only for a partial ranking of probability distributions. For example, two very different distributions - one having more inequality amongst poor, the other more inequality against the rich - can have the same index value.

When the distributional dominance fails, welfare-based measures, such as Atkinson Indexes, may provide for a \emph{complete} ranking among alternative distributions at the expense of more stringent assumptions as to how to represent social welfare \citep{Bellu2006}.  %leading to weaker outcome robustness 
Atkinson index has support [0,1] and 
is defined as
\begin{align*}
{\displaystyle \theta_A(\varepsilon)={\begin{cases}1-{\frac {1}{\hat{\mu }}}\left({\frac {1}{n_{iid}}}\sum _{i \in s_{iid}} y_{i}^{1-\varepsilon }\right)^{1/(1-\varepsilon )}&{\mbox{for}}\  \varepsilon \neq 1\\1-{\frac {1}{\hat{\mu} }}\left(\prod _{i \in s_{iid}} y_i\right)^{1/n_{iid}}&{\mbox{for}}\ \varepsilon =1.\end{cases}}}
\end{align*}
 The parameter $\varepsilon$ expresses the level of inequality aversion: as $\varepsilon$ increases, the index becomes more sensitive to changes at the lower end of the income distribution. 
 
 Besides, an additive decomposable family of inequality measures is the Generalized Entropy class. As opposed to the measures seen before, this class has the advantage of being strongly transfer-sensitive, meaning that it reacts to transfers depending on donor and recipient income levels. It is based on the concept of entropy which, when applied to income distributions, has meaning of deviations from perfect equality:
\begin{align*}
 \theta_{GE}(\alpha )={\begin{cases}{\frac  {1}{n_{iid}\alpha (\alpha -1)}}\sum_{i \in s_{iid}} \left[\left({\frac  {y_{i}}{\hat{\mu}}}\right)^{\alpha }-1\right]&\alpha \neq 0,1,\\{\frac  {1}{n_{iid}}}\sum _{i \in s_{iid}}{\frac  {y_{{i}}}{\hat{\mu}}}\ln {\frac  {y_{{i}}}{\hat{\mu}}}&\alpha \rightarrow1,\\-{\frac  {1}{n_{iid}}}\sum_{i \in s_{iid}}\ln {\frac  {y_{{i}}}{\hat{\mu}}}&\alpha \rightarrow 0.\end{cases}}
\end{align*}
The parameter $\alpha$ sets the sensitivity of the index: a large $\alpha$ induces the index to be more sensitive to the upper tail, and vice versa a small $\alpha$ to the lower tail. $\theta_{GE}(0)$ is the Mean Log Deviation, while $\theta_{GE}(1)$ is the well known Theil index. Atkinson and Generalized Entropy are two interrelated parametric families of measures, as a transformation of the Atkinson Index is a member of the GE class:
\begin{align*}
\theta_A(\varepsilon)=1-[\varepsilon(\varepsilon-1)\cdot \theta_{GE}(1-\varepsilon)+1]^{1/(1-\varepsilon)}.
\end{align*}
In this paper, we consider the estimation of both classes separately, since common parameter values used in one family do not correspond deterministically to parameter values commonly used for the other family.
Lastly, we consider the coefficient of variation (CV) as an inequality measure, being linked with a member of the GE family namely $\theta_{GE}(2)=$CV$^2/2$. Its square has been used in some income distribution analyses, including \citet{organisation2011divided}, even though it seems to be very sensitive to top outliers \citep{Atkinson2015}.

\section{Bias Correction Proposal}
\label{biasresults}

The bias of inequality estimators in small samples can be due to the structure of inequality measures as a non-linear function of estimators. The bias can be either positive or negative, depending on the characteristics of the reference variable distribution, except for the Mean Log Deviation which has a structurally negative bias as shown further on in this section. Among the measures with non-predictable bias direction, \citet{Breunig2001} shows that the bias of CV and GE$(2)$ is negatively related to the skewness of income distribution. This aspect could be analyzed in-depth by imposing a distributional assumption on the income variable, this is beyond the scope of this paper. For GE and Atkinson measures, the limiting behaviour of their bias is described in the following proposition.

\begin{proposition} 
For the measures belonging to the GE and Atkinson families, the expectation of their sample estimator $\hat{\theta}$, considering its true population value as $\theta$, can be expressed as:
 \begin{align*} \EX[\hat{\theta}]=\theta+O\bigg(\frac{1}{n_{iid}}\bigg),
  \end{align*}
  with $n_{iid}$ denoting the sample size in the $iid$ case.
\label{prop1}
%\vspace{mm}
\end{proposition} 

\begin{proof} In appendix.
\end{proof}

We are interested in a variety of non-linear functions of income values as inequality measures are. Let denote with $s$ a sample of size $n$, drawn using a complex sampling design, with $p(s)$ the probability of selecting the particular sample $s \subset \mathcal{U}$ out of the set of all possible samples $\mathcal{Q}$, thus $p(s)\geq 0$ and $\sum_{s\in  \mathcal{Q}} p(s) = 1$. The inclusion probability of unit $k$ is denoted with $\pi_k$, being $\pi_{k}=\sum_{s \in \mathcal{Q}_k} p(s)$ with $\mathcal{Q}_k$ the set of all possible samples including unit $k$.

We consider the generic inequality measure written as a function of the mean $\mu$ and $\gamma=\EX[g(y)]$, with $g(\cdot)$ a generic monotone transformation of the income variable. The population value for the generic inequality measure is 
\begin{align*}
\theta=f(\mu, \gamma),
\end{align*}
with $f(\cdot)$ a twice-differentiable function.
The related estimator in our complex survey framework is 
$\hat{\theta}=f(\hat{\mu}, \hat{\gamma})$
in which Horvitz-Thompson estimators of $\mu$ and $\gamma$ are plugged in, i.e.
\begin{align*}
\hat{\mu}=\frac{\sum_{i \in s} w_i y_i}{N}
\quad \text{and} \quad
\hat{\gamma}=\frac{\sum_{i \in s} w_i g(y_i, \boldsymbol w)}{N},
\end{align*}
where $\boldsymbol w=(w_1, \dots, w_n)=(1/\pi_1, \dots, 1/\pi_n)$ or a treated and calibrated version of it and $N$ is the population size. Note that the results of this section also hold for Hájek type estimators, i.e. with denominator $\hat{N}=\sum_{i=1}^n w_i$, since it is approximately unbiased \citep[pg. 182]{sarndal2003model}. \citet{kakwani1990large} uses a similar approach to express inequality indices to derive their asymptotic standard error. 
 By simply applying a second-order Taylor's series expansion of the sample estimator around the population values and evaluating its expected value, the bias can be expressed as
\begin{align}
\EX[\hat{\theta}-\theta]=&\frac{\partial f(\gamma, \mu)}{\partial \gamma} \EX[\hat{\gamma}-\gamma]+\frac{1}{2} \frac{\partial^2 f(\gamma, \mu)}{\partial \gamma^2} (\mathbb{V}[\hat{\gamma}]+ \EX^2[\hat{\gamma}-\gamma])+ \nonumber \\&+\frac{\partial^2 f(\gamma, \mu)}{\partial \gamma \partial \mu} (Cov[\hat{\gamma}, \hat{\mu}]-\mu  \EX[\hat{\gamma}-\gamma])+ \frac{1}{2} \frac{\partial^2 f(\gamma, \mu)}{\partial \mu^2}\mathbb{V}[\hat{\mu}]+O(n^{-2}),
\label{bias1}
\end{align}
 notice that $\hat{\mu}$ is unbiased. 

In Table \ref{tab_bias}, we detail the survey estimators for each inequality measure and their bias formulation based on Equation \ref{bias1} along with all relevant quantities. 
The complex survey estimators of Atkinson and Generalized Entropy measures come from \citet{Biewen2006}, while as for the Gini index, we employ the alternative formulation defined by \citet{sen1997economic}
and the complex survey estimator proposed by \citet{Langel2013}.
Let denote with $\sqrt{n/(n-1)}$ the standard bias-correction adjustment for the weighted variance;  $F(\cdot)$ denotes the cumulative distribution function of the variable of interest and lastly consider $\hat{N}_i=\sum_{k \in s} w_k \mathds{1}(n_k \leq n_i)$. The notation $\mathds{1}(A)$ defines an indicator function, assuming value 1 if $A$ is observed and 0 otherwise.

 Note that the bias formulas of Table \ref{tab_bias} 
 can also be reached differently, namely by applying the linearization proposed by \citet{graf2011use} and extended by \citet{vallee2019linearisation}, as made explicit in the Appendix. The Graf's methodology requires a separate derivation for each measure. In contrast, Equation \ref{bias1} defines a general formulation of the bias which applies to the entire set of considered measures, isolating its components and easing a general interpretation. 
 
Let us denote the Gini index estimator with $\hat{\theta}_G$, its approximate bias in small samples is
\begin{align}
\EX[\hat{\theta}_G-\theta_G]& \approx \frac{2}{\mu} \EX[\hat{\gamma}-\gamma] + \frac{2\gamma}{\mu^3} \mathbb{V}[\hat{\mu}]- \frac{2}{\mu^2} (Cov[\hat{\mu}, \hat{\gamma}]-\mu \EX[\hat{\gamma}-\gamma]) \label{biasgini} \\
&=\frac{4}{\mu} \EX[\hat{\gamma}-\gamma] + \frac{2\gamma}{\mu^3} \mathbb{V}[\hat{\mu}]- \frac{2}{\mu^2} Cov[\hat{\mu}, \hat{\gamma}], \nonumber
\end{align}
with $\gamma$ and $\hat{\gamma}$ as defined in Table \ref{tab_bias} and $\theta_G$ denoting the true value.
The derivation of the approximate bias related to the weighted estimator $\hat{\gamma}$ is not trivial. As explained by \citet{Langel2013}, its numerator is not composed of two simple sums. Indeed the quantity $\hat{N}_k$, an estimator of the rank of unit $k$, is random since its value depends on the selected sample.  One solution is to consider the approximate bias of the corresponding $iid$ estimator, i.e. $\EX[\hat{\gamma}-\gamma] =-1/n ( \gamma- \mu/2)$ as derived by \citet{Davidson2009}, so that: 
\begin{align}
\EX[\hat{\theta}_G-\theta_G] =\frac{-2\theta_G}{n} + \frac{2\gamma}{\mu^3} \mathbb{V}(\hat{\mu})- \frac{2}{\mu^2} Cov(\hat{\mu}, \hat{\gamma}).
\label{ginis}
\end{align}
This correction is in line with \citet{Davidson2009} and \citet{Fabrizi2016} proposals. However these are based on a first-order Taylor's expansions and thus limited to the first term of the right-hand side equation (\ref{biasgini}), ours extends it to a second-order expansion. This translates into the fact that, while \citet{jasso1979gini}, \citet{Deltas2003} and \citet{Davidson2009} proposals identify the adjusted Gini in $iid$ context as $n(n-1)^{-1}\hat{\theta}_G$, our correction reconsiders the shape of the adjusted estimator with a further order of approximation as
\begin{align}
\frac{n}{n-2}(\hat{\theta}_G-a), 
\label{shapegini}
 \end{align}
 with $a$ equals the sum of the second and third terms of (\ref{ginis}).  
 
As clear from Table \ref{tab_bias}, the bias correction of GE(2) does not include the coefficient of skewness of the income distribution, as shown by \citet{Breunig2001}. A reliable estimation of that quantity, while being straightforward in the $iid$ case, appears cumbersome in the case of weighted data being defined on a discrete grid of values. This leads to the non-applicability of \citet{Breunig2001} result in our case.

\begin{sidewaystable}[]
\setlength{\extrarowheight}{1.5em}
\setlength{\extrarowheight}{2em}
\begin{tabular}{ l l l l l l }
 Measure &  \textbf{$\gamma=\EX[ g(y)]$}  &  Design Estimator &  $\hat{\gamma}$ &  $ f(\hat{\mu},\hat{\gamma})$ &  Approximate Bias \\ 
\toprule
Gini & $\EX[y\cdot F(y)]$ &  $\frac{2 \sum_{i \in s} w_i y_i (\hat{N_i}-w_i/2) }{N^2\hat{\mu}}-1$ &
$\frac{ \sum_{i \in s} w_i y_i (\hat{N_i}-\frac{w_i}{2}) }{N^2}$ &
$\frac{2\hat{\gamma}}{\hat{\mu}} -1$ & \makecell[l]{$\frac{4}{\mu} \EX[\hat{\gamma}-\gamma] + \frac{2\gamma}{\mu^3} \mathbb{V}[\hat{\mu}]- \frac{2}{\mu^2} Cov[\hat{\mu}, \hat{\gamma}]$}\\
\hline
\makecell[l]{GE($\alpha)$ \\ $\alpha \neq 0,1$} &
$\EX[y^\alpha]$ & 
$\frac{n(n-1)^{-1}}{\alpha(\alpha-1)} \bigg[\frac{\sum_{i \in s} w_i y_i^{\alpha}}{N\hat{\mu}^\alpha}-1 \bigg]$ & $\frac{\sum_{i \in s} w_i y_i^\alpha}{ N}$ &  
$\frac{n(n-1)^{-1}}{\alpha(\alpha-1)}  \bigg[\frac{\hat{\gamma}}{\hat{\mu}^\alpha}-1 \bigg]$ &
\makecell[l]{$\frac{n(n-1)^{-1}}{\mu^{\alpha+1}(\alpha-1)}\bigg[ \frac{\gamma(\alpha+1)}{2\mu} \mathbb{V}[\hat{\mu}] -Cov[\hat{\gamma},\hat{\mu}]\bigg]$}\\
GE($0$) & $\EX[\log y]$ & 
$\frac{1}{N} \sum_{i \in s} w_i \log \frac{\hat{\mu}}{y_i}$ & $\frac{ \sum_{i \in s} w_i \log y_i}{N} $ & $\log(\hat{\mu})- \hat{\gamma}$ &
\makecell[l]{$ -\frac{1}{2\mu^{2}} \mathbb{V}[\hat{\mu}]$}\\
GE(1) & $\EX[y(\log y)]$ & 
$\frac{1}{N} \sum_{i \in s} w_i  \frac{y_i}{\hat{\mu}} \log \frac{y_i}{\hat{\mu}}$ &
$\frac{\sum_{i \in s} w_i y_i \log y_i}{N}$ &
$\frac{\hat{\gamma}}{\hat{\mu}}-\log(\hat{\mu})$ & \makecell[l]{$\bigg[\frac{\gamma}{\mu^{3}}+\frac{1}{2\mu^{2}} \bigg]\mathbb{V}[\hat{\mu}]-\frac{1}{\mu^{2}}Cov[\hat{\mu}, \hat{\gamma}]$} \\
\hline
\makecell[l]{A($\varepsilon)$ \\$\varepsilon \neq 1$ }& 
$\EX[y^{1-\varepsilon}]$&
$1-\frac{1}{\hat{\mu}} \bigg[ \frac{1}{N} \sum_{i \in s} w_i y_i^{1-\varepsilon}\bigg]^{\frac{1}{1-\varepsilon}}$ &
$\frac{\sum_{i \in s} w_i y_i^{1-\varepsilon}}{  N}$ &
$1-\frac{\hat{\gamma}^{\frac{1}{1-\varepsilon}}}{\hat{\mu}}$ &
\makecell[l]{$\frac{\gamma^{\frac{\varepsilon}{1-\varepsilon}}}{\mu}\bigg[\frac{Cov[\hat{\gamma}, \hat{\mu}]}{\mu(1-\varepsilon)} -\frac{\gamma}{\mu^{2}}\mathbb{V}[\hat{\mu}]-\frac{\varepsilon}{2\gamma(1-\varepsilon)^2} \mathbb{V}[\hat{\gamma}]\bigg]$}\\
A(1) &
 $\EX[\log y]$ &
$1- \frac{1}{\hat{\mu}} \prod_{i \in s} y_i^{w_i/N}$ & 
$\frac{\sum_{i \in s} w_i \log y_i}{N}$ &
$1-\frac{\exp \{ {\hat{\gamma}} \} }{\hat{\mu}}$ &
$\makecell[l]{\frac{\exp \{ \gamma \}}{\mu^2} \bigg[
Cov[\hat{\gamma}, \hat{\mu}]-\frac{\mu}{2}\mathbb{V}[\hat{\gamma}]- \frac{1}{\mu}\mathbb{V}[\hat{\mu}]\bigg]}$\\
\hline
CV & $\EX[y^2]$ &
$\bigg[ \frac{n}{N(n-1)} \sum_{i \in s} w_i \frac{y_i^2}{\hat{\mu}^2}-1 \bigg]^{\frac{1}{2}}$ & 
$\frac{\sum_{i \in s} w_i y_i^2}{N}$ &
$\bigg[\frac{n}{n-1} \frac{\hat{\gamma}-\hat{\mu}^2}{\hat{\mu}^2}\bigg]^{\frac{1}{2}}$ &
\makecell[l]{$\sqrt{\frac{n}{n-1}}  \frac{1}{\mu^3}   \bigg(\frac{\gamma}{\mu^2}-1\bigg)^{-\frac{3}{2}}\bigg[ 
\mathbb{V}[\hat{\mu}] \frac{\gamma}{2\mu} \bigg( \frac{2 \gamma}{\mu^2}-3\bigg)-$ \\$- Cov[\hat{\gamma}, \hat{\mu}] \bigg( \frac{\gamma}{2\mu^2}-1\bigg)- \frac{1}{8\mu} \mathbb{V}[\hat{\gamma}]\bigg]$}\\
\bottomrule
\end{tabular}
\caption{Relevant quantities for each measure including the approximate bias.}
\label{tab_bias}
\end{sidewaystable}

\section{Bias Estimation}
\label{biasest}

In this section, we detail the estimation of the approximate bias defined in Table \ref{tab_bias} for each measure. Such estimation is not trivial considering that the mentioned expressions depend on design variances and covariances $\mathbb{V}[\hat{\mu}]$, $\mathbb{V}[\hat{\gamma}]$ and $Cov[\hat{\mu}, \hat{\gamma}]$. We consider a complex survey design involving stratification and multi-stage selection, with both Self-Representing (SR), i.e. included at the first stage with probability one, and Non-Self-Representing (NSR) strata. This design is consistent with the majority of income survey designs and, in general, with official statistics household surveys.

We define an unbiased estimator for the variance of Horvitz-Thompson estimators, such as $\hat{\mu}$, when $w_i=1/\pi_i$ as 
\begin{align*}
\hat{\mathbb{V}}[\hat{\mu}]= \frac{1}{N^2} \bigg( \sum_{i \in s} y_i^2 \frac{1-\pi_i}{\pi_i^2} +  \sum_{i \in s}   \sum_{k \in s, i \neq k}  y_iy_k\frac{\pi_{ik}-\pi_i \pi_k}{\pi_{ik}} \bigg),
\end{align*}
with $\pi_{i k}$, $\forall  i, k \in \mathcal{U}, i\neq k$ denoting the second-order inclusion probabilities i.e. the probability that the sample includes both $i$-th and $k$-th units \citep[pg. 30]{arnad2017}.
However generally (a) $w_i \neq 1/\pi_i$ and (b) $\pi_{ik}$, $\forall  i, k \in \mathcal{U}, i \neq k$ are difficult to calculate under complex sampling designs. 
	
  Therefore, the variance estimator to be considered constitutes an approximation that relies on simplified assumptions. Firstly, we assume that Primary Sampling Units (PSU) are sampled with replacement, and secondly, we reduce multi-stage sampling into a single-stage process by relying on the Ultimate Clusters technique \citep{kalton1979ultimate}. Moreover, we take into account the hybrid nature of the probability scheme, blending a variance estimator for stratified design associated with the SR strata, including a finite population correction factor, and a typical Ultimate Cluster variance estimator for multi-stage schemes associated with the NSR strata. The latter one is widely used in official statistics, see \citet{Osier2013} for Eurostat procedures. 
Without loss of generality, let us consider a two-stage scheme, where $\hat{\mu}=\sum_{h}\sum_{d}\sum_{i} w_{hdi} y_{hdi}/N$ is a linear estimator of $\mu$, with $h$ the stratum indicator, $d$ the Primary Sampling Unit (PSU) indicator and $i$ the secondary sampling unit (household) indicator. Its variance estimator 
is as follows:
 \vspace{-2mm}
\begin{align}
\hat{\mathbb{V}}[\hat{\mu}]& =\sum_{h=1}^{H_{SR}} \mathbb{V}[\hat{\mu}_h]+\sum_{h=1}^{H_{NSR}} \mathbb{V}[\hat{\mu}_h]\nonumber\\
&= \sum_{h=1}^{H_{SR}} M_h^2 (1-f_h) \frac{s_{h}^2}{m_h} +\sum_{h=1}^{H_{NSR}} n_h   s_{\hat{\mu}_h}^2 \label{eurostat}\\
&=\sum_{h=1}^{H_{SR}} M_h \frac{M_h-m_h}{m_h(m_h-1)} \sum_{i=1}^{m_h} (y_{hi}- \bar{y}_{h} )^2+\sum_{h=1}^{H_{NSR}} \frac{n_h}{n_h-1} \sum_{d=1}^{n_h} (\hat{\mu}_{hd}- \bar{\mu}_{h})^2,\nonumber
\end{align}
with $H_{SR}$ self-representative and $H_{NSR}$ non self-representative strata, $M_h$ the number of resident households in strata $h$, $m_h$ the number of sample households in strata $h$, $f_h=m_h/M_h$ a finite population correction factor, $n_h$ the number of PSUs in strata $h$.
Consider, moreover, that $\bar{y}_{h} = \sum_{i=1}^{m_h} y_{hi}/ m_h$, $\hat{\mu}_{hd}= \sum_{i=1}^{m_d} w_{hdi}y_{hdi}/N$ with $i$ denoting the household label and $m_d$ the number of sample households in PSU $d$, lastly $\bar{\mu}_{h}=\sum_{d=1}^{n_h}\hat{\mu}_{hd}/n_h$,  with $n_h$ being the number of PSU in stratum $h$. 
Obviously, if $n_h = 1$ for some strata, the estimator (\ref{eurostat}) cannot be used. A solution is to collapse strata to create “pseudo-strata” so that each pseudo-stratum has at least two PSUs. Common practice is to collapse a stratum with another one that is similar with respect to some survey target variables  \citep{rust1987strategies}.

An estimator of $\mathbb{V}[\hat{\gamma}]$ can be obtained by adopting the same strategy used for $\mathbb{V}[\hat{\mu}]$ in (\ref{eurostat}). Whereas, regarding the estimation of the design covariance, consider that 
\begin{align*}
Cov[\hat{\gamma}, \hat{\mu}]= \frac{1}{2}\bigg( \mathbb{V}[\hat{\gamma} +\hat{\mu}]-
\mathbb{V}[\hat{\gamma}]-\mathbb{V}[\hat{\mu}] \bigg).
\end{align*}

Thus, a possible estimator $\hat{Cov}[\hat{\gamma}, \hat{\mu}]$ would be simply obtained by plugging in the variance estimators previously mentioned, while $\mathbb{V}[\hat{\gamma} +\hat{\mu}]$ is estimated by considering $\hat{\gamma} +\hat{\mu}=\sum_{i \in s} w_i (g(y_i)+y_i) /N$. 
The estimation procedure is completed by replacing $\mu$ and $\gamma$ with $\hat{\mu}$ and $\hat{\gamma}$.

The Gini index estimator differs from the other indexes since $\hat{\gamma}$ is a non-linear statistic. Thus, a linearization of $\hat{\gamma}$ is needed to make it tractable and carry on variance estimation with the procedure described above. We consider again the linearization proposed by \citet{graf2011use} with the practical adaptation of \cite{Graf2014a} for inequality estimators.
In such adaptation, the linearized variable is merely a function of the partial derivatives with respect to the weights, that in the case of $\hat{\gamma}$ defined for Gini index in Table \ref{tab_bias} is
\begin{align*}
v_k= \frac{\partial \hat{\gamma}}{\partial w_k}=\frac{1}{N^2} \bigg[y_k (\hat{N}_k-w_k) + \sum_{i \in \mathcal{S}_k}  w_i y_i \bigg],
\end{align*}
for a generic unit $k$ where $\mathcal{S}_k=\lbrace i \in s, n_i > n_k \rbrace$. In this way, the estimator can be re-expressed through a linear approximation, namely $\hat{\gamma} \approx \sum_{i \in s} w_i v_i$, and it becomes possible to perform variance estimation of linear statistics.

\section{Design-Based Simulation}
\label{simulationsection}

A design-based simulation study has been conducted to evaluate our bias correction proposal. In this simulation, the cross-section Italian EU-SILC sample (2017 wave) has been assumed as pseudo-population and the 21 NUTS-2 regions have been considered as target domains. The study is based on real income data, in order to check whether this specific framework works with close-to-reality data, affected by peculiar problems (e.g. extreme values, skewness). 

For comparison purposes, two simulation scenarios have been carried out. In the first one, the original income data are employed as pseudo-population. In the second one, an extreme values treatment is performed concerning both upper and lower tails, to circumvent non-robustness problems. The resulting dataset is specified as an alternative pseudo-population. We compare the results obtained after the treatment with the ones before treatment to isolate the effect of extreme values when evaluating bias-correction performances (Table \ref{Tab:Tcr}). %on the performance of the corrections proposed

The issue of robust estimation of economic indicators through an extreme values treatment in the upper tail of income distribution is well-established in the literature. See \citet{Brzezinski2016} for a review and \citet{Alfons2013} for a suitable specification for survey data. On the contrary, the issue of treatment of extreme values in the lower tail of income distribution appears less established \citep{Kerm2007, masseran2019power}. 
Concerning the upper tail, we operated a semi-parametric Pareto-tail modelling procedure using the Probability Integral Transform Statistic Estimator (PITSE) proposed by \citet{Finkelstein2006}, which blends very good performances in small samples and fast computational implementation, as suggested by \citet{Brzezinski2016}. As regards the lower tail, we used an inverse Pareto modification of the PITSE estimator, suggested by \citet{masseran2019power}.
In our simulations, the treatment has been done at a regional level to the original EU-SILC sample and the detection of extreme values has been carried out following \citet{MohdSafari2018} by using the Generalized Boxplot procedure. 
We expect that, when outlying observations are representative, such treatment would highly bias the outcome and thus we do not recommend it.

From both pseudo-populations, we repeatedly select 1,000 two-stage stratified samples, mimicking the sampling strategy adopted in the survey itself: in the first stage, SR strata are always included in the sample, while a stratified sample of PSU in NSR strata is selected; in the second stage, a systematic sample of households is drawn from each PSU included at the first stage. 
We repeated the drawing for both scenarios involving different sampling rates, 1.5\% and 3\% respectively. The Relative Bias (RB) and the Absolute Relative Error (ARE) in percentage have been calculated for each region $r$ using the 1,000 iterations as:
\begin{align*}
\texttt{RB}_r&=  \frac{1}{1,000} \sum_{p=1}^{1,000}\bigg( \frac{\hat{\theta}_{p,r}}{\theta_r} -1 \bigg), \\
\texttt{ARE}_r&=\frac{1}{1,000}  \bigg( \sum_{p=1}^{1,000}   \bigg|  \frac{\hat{\theta}_{p,r}}{\theta_r} -1 \bigg| \bigg),
\end{align*}
\noindent where $\theta_r$ is the population value for region $r$ and $\hat{\theta}_{p,r}$ its estimate at iteration $p$.
In our simulation setting, the regional sample sizes range from 6 to 96 individuals (from 6 to 32 households) on average over the simulated samples for the 1.5\% sampling rate, and from 11 to 196 individuals (10 to 74 households) for the 3\% sampling rate.

\begin{figure}
    \centering
    \includegraphics[width=\textwidth]{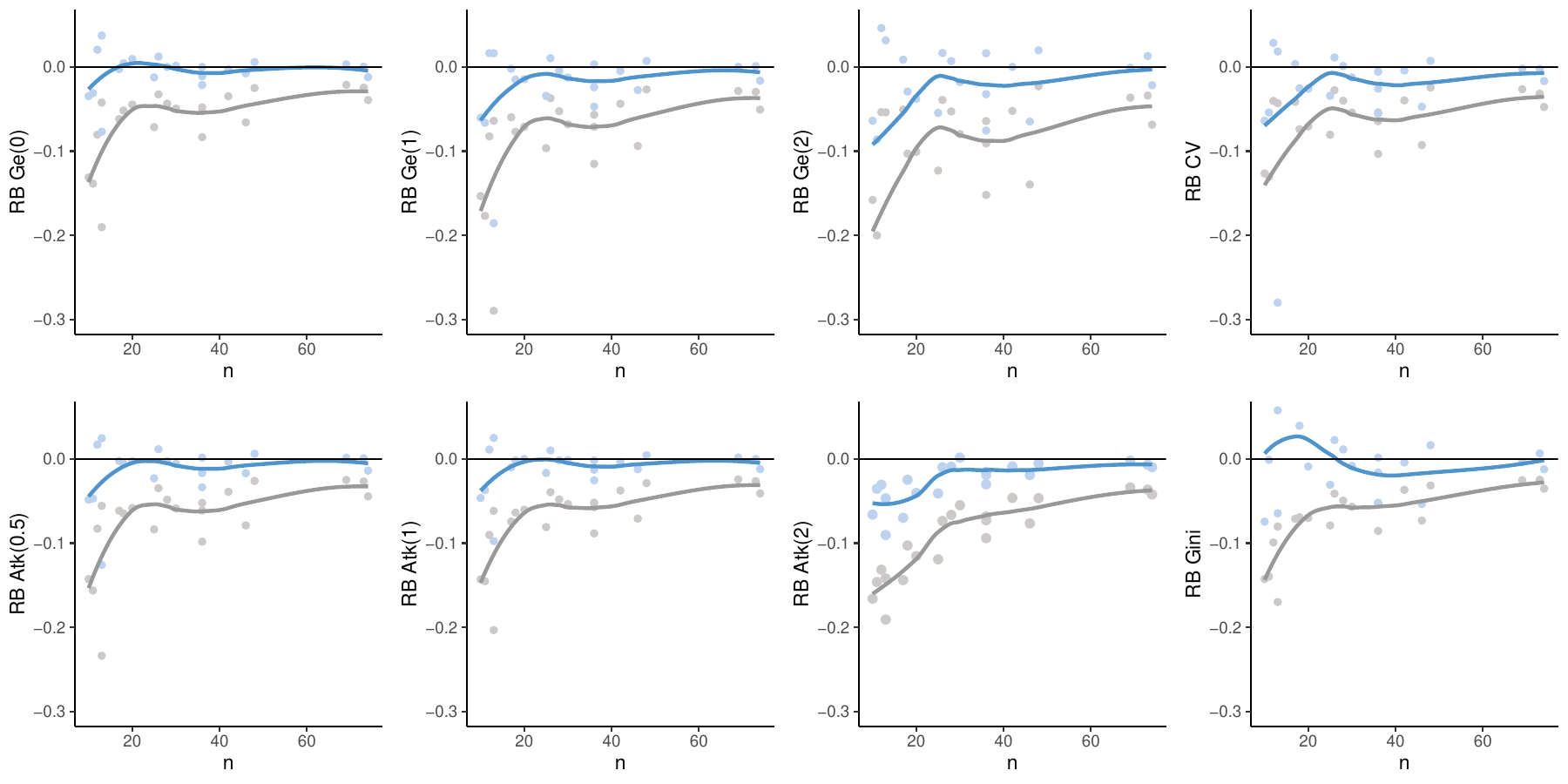}
  \caption{Relative Bias of non-corrected measures (grey line), and of corrected measures (blue line) in 3\% samples after extreme value treatment versus the (average) sample size.}
  \label{plotbias}
\end{figure}

\begin{table}
\centering
\begin{tabular}{rrrrrrrrrr}
 \toprule
 && \hspace{3mm}CV & GE(0) & GE(1) & GE(2) &A(0.5) & A(1) & A(2) & Gini\\ 
   \hline
   &&&\\
   \multicolumn{4}{c}{with extreme values treatment}&&&&&&\\
  \hline
   &&&\\
 1.5\% &&&&&&&&&\\
   \hline
&$\bar{\texttt{RB}}$ &-11.9& -13.9 & -16.0 &-17.6&-14.9& -15.0 & -19.0 &  -14.6\\ 
   $\hat{\theta}\quad$& $\bar{\texttt{ARE}}$ &25.8& 44.1 & 42.9 &47.9& 41.9 &40.9 & 38.7 & 24.5\\ 
  &&&\\
&$\bar{\texttt{RB}}$ &-5.6& -4.1 & -6.8 &-9.0& -5.4&-5.2 & -9.4 & 0.5\\ 
   $\hat{\theta}_{corr}$ &  $\bar{\texttt{ARE}}$ &25.9& 46.4 & 44.9 &50.2&  43.9 &42.8 & 39.6 &34.3\\ 
      \hline
 &&&\\
 3.0\%&&&&&&&&& \\
    \hline
 &$\bar{\texttt{RB}}$ &-7.4& -6.6 & -8.6 & -10.6&-7.6&-7.3 & -9.7 & -7.2\\ 
   $\hat{\theta}\quad$  &$\bar{\texttt{ARE}}$ &19.8& 32.0 & 32.0&38.0& 30.6 & 29.6 & 28.2 & 16.6\\ 
 &&&\\
  & $\bar{\texttt{RB}}$ &-2.8& -0.6 & -2.3 &-3.6 & -1.5 &-1.2& -2.9 & 0.3\\ 
&$\bar{\texttt{RB}}$ ($n\geq20$)& -1.4 & -0.4 & -1.2 & -1.6 &  -0.8 &-0.6 & -1.4 &-0.9 \\ 
   $\hat{\theta}_{corr}$ & $\bar{\texttt{ARE}}$ &20.4& 33.4 & 33.7 &40.6& 32.0 & 30.4 & 29.6 &19.7\\
    \bottomrule
   &&&\\
      \multicolumn{4}{c}{without extreme values treatment}\\
  \hline
  &&&\\
   1.5\% &&&&&&&&&\\
       \hline
&$\bar{\texttt{RB}}$ &-18.2 & -12.7 & -17.5 & -23.3&-15.3 & -15.6& -48.0 &  -14.9 \\ 
  $\hat{\theta}\quad$& $\bar{\texttt{ARE}}$ & 30.0 & 52.9 & 46.4 & 53.5 &45.9  & 47.4 & 56.8 &  25.5 \\ 
 &&&\\
  &$\bar{\texttt{RB}}$ & -12.1 & -3.9 & -8.7 & -15.0 &  -6.3 & -5.9 & -41.6 &0.04 \\ 
  $\hat{\theta}_{corr}$& $\bar{\texttt{ARE}}$ & 29.3 & 54.4 & 48.0 & 55.7 & 47.5 & 49.4 & 54.6 & 34.7 \\ 
      \hline
 &&&\\
   3.0\%&&&&&&&&& \\    
       \hline 
     &$\bar{\texttt{RB}}$ & -12.7 & -6.8 & -10.5 & -15.8 & -8.7 &-8.4 & -38.1 &  -7.9 \\ 
$\hat{\theta}\quad$&  $\bar{\texttt{ARE}}$ &24.5 & 39.4 & 36.0 & 46.2 &34.3 & 35.6 & 49.0 &  17.7\\ 
 &&&\\
& $\bar{\texttt{RB}}$ & -7.8 & -1.2 & -3.9 & -8.0 &-2.5 &-2.0 & -32.4 &   0.06 \\ 
&$\bar{\texttt{RB}}$ ($n\geq20$)& -8.3 & -1.2&  -3.6&  -8.7&-2.3&  -1.7& -30.0&   -1.2\\
 $\hat{\theta}_{corr}$   & $\bar{\texttt{ARE}}$ & 24.8 & 40.4  & 37.9 & 49.4 &35.7& 37.0 & 48.2 &  20.9\\ 
  \bottomrule
\end{tabular}
\caption{Percentage RB and ARE averaged on the 21 regions for each inequality estimator and scenario.}
\label{Tab:Tcr}
\end{table}

Concerning the treated pseudo-population scenario, Figure \ref{plotbias} illustrates the relative bias for each domain of non-corrected measures (grey line) and of corrected measures (blue line) in 3\% samples versus the (average) sample size. The negative relation between sample size and average relative bias is clear for both the survey estimator $\hat{\theta}$ and the bias-corrected estimator $\hat{\theta}_{corr}$. This confirms the nature of the bias as a small sample bias and shows the effectiveness of the correction, even if based on a large-n approximation as the Taylor's expansion. The bias reduction is noticeable for all measures, leading to slightly biased estimates depending on the measure. Notice that the bias correction works well for measures that are not particularly sensitive to extreme observations such as the Gini index, GE$(0)$, Atk$(0.5)$ and Atk$(1)$. In the case of CV and GE$(2)$, the correction provides good results, but it seems, however, not to capture all the bias components. This may confirm the results of \citet{Breunig2001}, suggesting that the coefficient of variation squared and GE(2) bias depends on the coefficient of skewness of the income distribution, not considered in our bias correction.

Bias and error averaged across all areas for each scenario, sampling rate and estimator are shown in Table \ref{Tab:Tcr}. By still focusing on treated population results, the correction induces a reduction of the RB spanning from 5\% (CV, 3\% rate) to 14\% (Gini, 1.5\% rate) approximately by considering both sampling rates.
When the sample size is greater than 20 individuals ($n \geq 20$), the bias-corrected estimators seem to be approximately unbiased.
Furthermore, it is important to note that the bias correction induces a slight but negligible error (ARE) increase for every measure, except for the Gini index which presents a relevant increase. This exception may be explained by the shape of the unbiased estimators, as described by (\ref{shapegini}), where a sum of estimators is multiplied by a factor  $n/(n-2)$, which inherently inflates the variance by its square.  

Let us focus on comparing the treated population scenario with the non-treated one. In the latter case, bias and error increase dramatically both for $\hat{\theta}$ and for $\hat{\theta}_{corr}$. In particular, the bias is great for some measures estimated on the non-treated scenario due to their non-robustness properties to extreme values. It is the case of Atk ($\varepsilon=2$), extremely sensitive to low-income values (under $100$ euro per year) which is -48\% biased on average for the scenario with the smallest sample sizes. Also, GE with $\alpha$ equal to 1 and 2 are highly sensitive to high-income values being -18\% and -23\% biased. However, the bias correction leads to a bias reduction comparable in magnitude to the one discussed for the treated pseudo-population; it seems not to change in magnitude with respect to the sample size and the presence of extreme values. % No relevant changes in the comparison of $\hat{\theta}$ and $\hat{\theta}_{corr}$ are recorded.

 To summarize, our results highlight that in the case of populations that are not affected by income extreme values, the bias correction may provide approximately unbiased estimates for a large class of measures at the expense of, in most cases, only a slight error increase. Vice versa, it might be necessary to restrict the attention to the most robust measures such as GE with $\alpha=0$, Atkinson index with $\varepsilon=1$ and Gini Index to obtain estimates affected by a negligible bias. Another important aspect to point out is that, in certain countries, the EU-SILC is based on registers that better capture top incomes, thus, a cross-country comparison of income inequality by effects on a tail-sensitive measure must be another reason for caution \citep{Atkinson2015}. Such results may constitute a reference when measuring inequality in small samples, however, since the simulation scenario uses specific data, reflections that have been drawn cannot be general or conclusive.

\section{A Small Area Estimation Exercise}
\label{smallareaexercise}

In the previous sections, we propose a method to correct the small sample bias of inequality estimators in complex surveys. Even if bias-corrected, such estimators are still unreliable due to the high variability induced by the small sample size: this means that estimates cannot be released or used for further inference. As a consequence, when measuring inequality at a fine-grained level, it becomes necessary to rely on Small Area Estimation (SAE) techniques. Such estimation techniques take advantage of available auxiliary information to produce estimates with acceptable uncertainty. More specifically, the model-based SAE techniques employ hierarchical models which can be defined both at area-level, linking area-defined survey estimates with areal covariates, or at unit (individual) level, linking individual income data with individual covariates. See \citet{tzavidis2018start} for an up-to-date review.

In this context, area-level models appear to be less demanding in terms of data requirements and enable the incorporation of design-based properties. Such models constitute a typical framework of application of our bias-correction proposal, as they assume the unbiasedness of survey estimators used as input. As a consequence, their applicability to the estimation of inequality measures is inevitably tied to a preliminary bias correction, in contrast with unit-level models that do not involve survey estimators.

In this section, we perform an SAE exercise by using the sample related to the first iteration of the simulation detailed in Section \ref{simulationsection} for the 3\% case. The purpose is not to propose a small area estimation strategy nor to provide a real application of inequality mapping, but rather to illustrate the framework of application of our bias-correction proposal and, especially, to underline the risk of avoiding bias-correction when estimating inequality in small domains. Such exercise is carried out by applying the Fay-Herriot model \citep{fay1979estimates}, a landmark model in the small area literature, implemented through the package \texttt{sae} \citep{molina2015sae} to both uncorrected and corrected survey estimators. The objective is to check whether the inclusion of biased or bias-corrected survey estimates in the model may lead to different results. From the whole set of inequality estimators considered in Sections \ref{biasresults}, \ref{biasest} and \ref{simulationsection}, we perform the exercise on the most popular ones: the Theil index (Generalized Entropy with $\alpha=1$), the Atkinson index with $\varepsilon=1$ and the Gini index. 

Specifically, let us consider $\hat{\theta}_1, \dots, \hat{\theta}_M$ as the set of survey estimators referring to a generic inequality measure in $M$ small areas, with corresponding population values $\theta_1, \dots, \theta_M$, and $\boldsymbol x_m$ the set of $p$ areal covariates for area $m$, $m=1, \dots, M$.
The classical area-level model is the Fay-Herriot  one, specified as follows:
\begin{align}
&\hat{\theta}_m \sim \mathcal{N}(\theta_m, D_m),\\
&\theta_m \sim \mathcal{N}(\boldsymbol x_m^T \boldsymbol \beta,  \sigma^2), \quad \quad  m=1, \dots, M.
\end{align}
\noindent where $D_m$ denotes the sampling variance of the survey estimator, usually assumed to be known to allow for identifiability, $\boldsymbol \beta$ the set of regression coefficients and $\sigma^2$ the model variance.
This clearly implies $\mathbb{E}(\hat{\theta}_m)=\theta_m$ $\forall m$, i.e. the unbiasedness of survey estimators. As a consequence, neglecting the bias correction of survey estimators effectively leads to model misspecification.

As mentioned above, the sampling variance is separately estimated from the data and given as input to the small area model. Since our exercise is merely illustrative, we adopt the Monte Carlo variances of the design-based simulation in Section \ref{simulationsection} as sampling variances and simulated covariates for both estimators.
However, in real applications, variance estimation is the crux of an SAE procedure. 
In the case of uncorrected inequality estimators, it may be easily carried out via linearization. Linearized variables for each measure could be derived consistently with \citet{Langel2013} for the Gini index and \citet{Biewen2006} for the Generalized Entropy and the Atkinson indexes. On the other hand, the variance of bias-corrected estimators adds a new level of complexity since the estimator formula is no longer the classical one. Indeed, it comprises a bias correction component that appears cumbersome to estimate via linearization since it is inherently a result of several linearizations. Therefore, in real applications, we recommend relying on resampling methods. An example of such a method is the design-aware bootstrap procedure developed by \citet{Fabrizi2011, fabrizi2020functional}. A comprehensive review of the use of bootstrap methods for survey data can be found in \citet{lahiri2003impact}.

\begin{figure}[h]
    \centering
    \includegraphics[width=\textwidth]{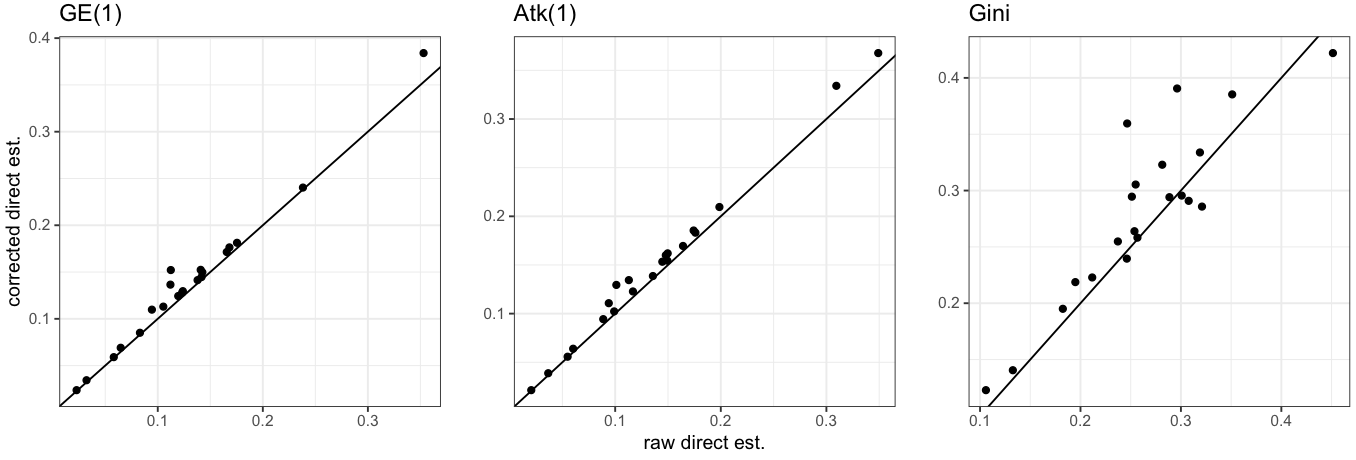}
  \caption{Bias corrected survey estimates versus uncorrected survey estimates. Bisector line in black. }
  \label{rawest}
\end{figure}

The comparison between uncorrected and corrected survey estimates for all three measures is displayed in Figure \ref{rawest}. Uncorrected estimates show lower values of inequality in comparison with the corrected ones for all areas and all the measures considered. This is in accordance with the underestimation highlighted by simulation results of Section \ref{simulationsection}. %Gini index results show greater dispersion than the other measures due to the instability induced by the $n/(n-2)$ term. 
The sampling coefficient of variations of both estimators are high, ranging from 0.24 to 3.48 for the Theil index, from 0.18 to 4.45 for the Atkinson index and, lastly from 0.11 to 0.92 for the Gini index depending on the area, with slightly higher values in case of corrected estimators as the bias correction induces mild variance inflation. Such values point out the need for SAE techniques.

\begin{figure}[h]
    \centering
    \includegraphics[width=\textwidth]{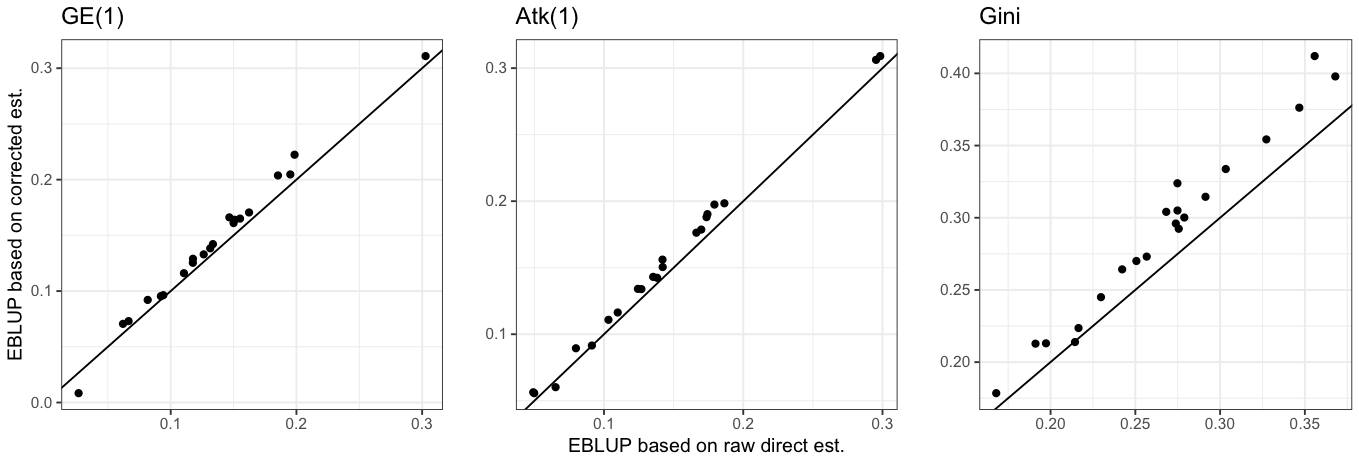}
  \caption{Model-based estimates based on bias-corrected survey estimates versus model-based estimates based on uncorrected survey estimates. Bisector line in black.}
   \label{modest}
\end{figure}

The model-based (or EBLUP, Empirical Best Linear Unbiased Predictor) estimates in both cases are compared in Figure \ref{modest}. The inequality levels estimated by the misspecified model are lower, resulting in a misleading inference. In particular, this is quite evident in the Gini index case, where the divergence seems to increase at increasing levels of inequality. When the same SAE procedure is applied for all iterations referred to the simulation of Section \ref{simulationsection}, a bias evaluation shows that the EBLUP estimated on uncorrected and corrected measures have an average RB of -17.9\% and -13.4\% respectively for the Generalized Entropy ($\alpha=1$); of -14.6\% and -9.4\% for the Atkinson index ($\varepsilon=1$) and of -7.7\% and -0.4\% for the Gini index. This confirms the risk of high underestimation of inequality when neglecting such an issue.

\begin{figure}[h]
    \centering
    \includegraphics[width=\linewidth]{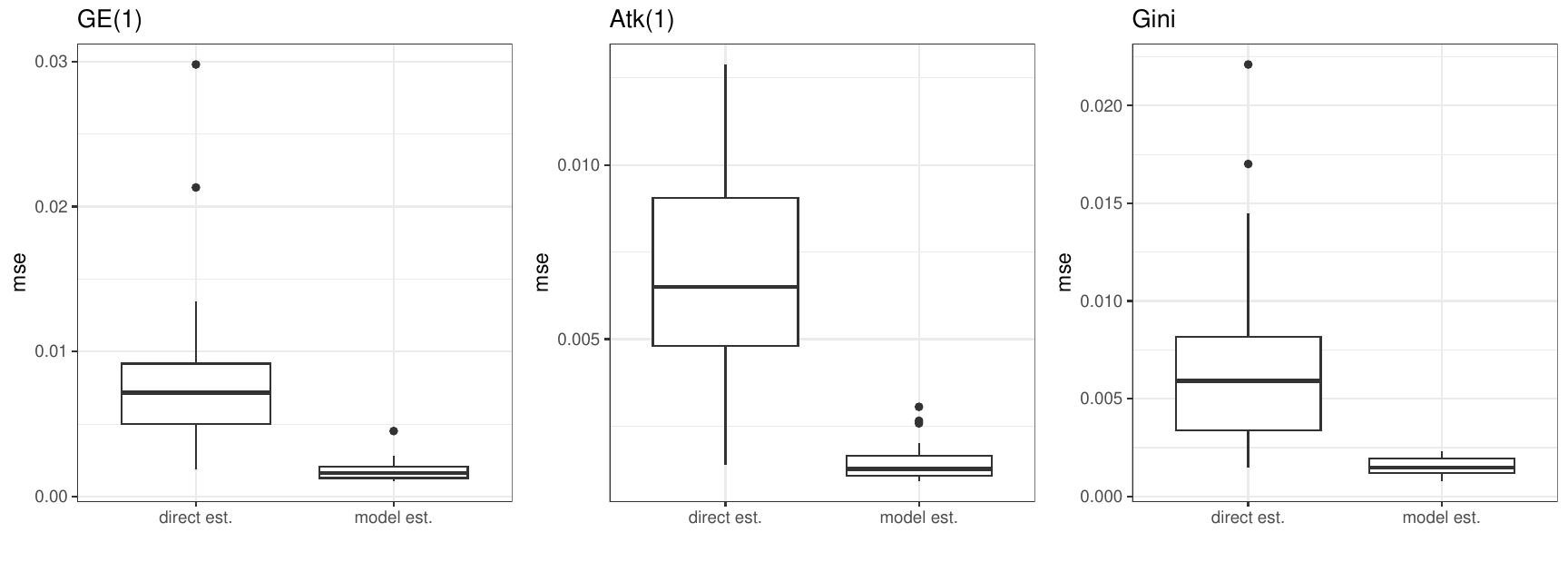}
  \caption{MSE of bias-corrected survey estimators versus their related model-based estimators.}
   \label{mse}
\end{figure}

By focusing only on EBLUP results based on corrected estimates, the decrease in terms of error induced by the model is depicted in Figure \ref{mse}. The reduction is relevant and testifies that the variance reduction procedure, put in place by the SAE model, is effective. As a consequence, such model-based estimates result to be reliable and ready to be used for further analysis or mapping.

%\vspace{5mm}

\section{Conclusions}
\label{conclusions}

A strategy based on Taylor's expansion has been proposed to correct the small sample bias of inequality estimators. The inequality measures considered are several, as the comparison of diverse measures may enable us to enlighten the specific point of view that each measure provides, like single tiles in a mosaic. Indeed, the well-known Gini and Theil indexes are widely applied in several fields for inequality and concentration estimation. 

A sensitivity analysis with respect to outliers  and a simulation study have been conducted to study the estimator behaviour to extreme values and the performance of the proposed correction. Results show that survey-based estimators may be biased in small samples, inducing an underestimation that is even greater in the case of populations affected by extreme values. Moreover, simulation results validate the correction proposal as effective, consistently reducing the bias and leading in some cases to approximately unbiased estimators. 

An underlined heterogeneity of sensitivities and bias is recorded across measures. As a consequence, our results may help in choosing the most suitable inequality measure depending on the context. 
The measures which are structurally more sensitive to extreme values appear to be more biased, in particular, GE with $\alpha=2$ and Atkinson with $\epsilon=2$. Therefore, in the case of samples without extreme income values, the bias correction may provide approximately unbiased estimates. On the other hand, if extreme values are observed, it becomes 
 necessary to focus on the most robust measures such as Mean Log Deviation, Atkinson index with $\varepsilon=1$ and Gini Index to be corrected. 

An illustrative small area application has been carried out to evaluate the effect of disregarding bias in a typical small-sized sample context. The results obtained show that neglecting it translates into a misleading inference and an inequality underestimation. This is particularly evident in the case of the popular Gini index.
In such an application, we use a basic area-level model, the Gaussian one. Indeed, the possibly not-Gaussian sampling distributions of inequality estimators and the unit-interval support of Gini and Atkinson estimators might urge a more refined model, which may lead to model-based estimators with increased performances: this suggests an interesting direction for future research. Further directions also include the extension of this framework to other widely used inequality measures, such as those based on quintiles and the development of a multivariate SAE framework. 

\section*{Acknowledgements}
The work of Silvia De Nicolò was partially funded by the ALMA IDEA 2022 grant (title: "Social exclusion and territorial disparities: poverty and inequality mapping through advanced methods of small area estimation", project J45F21002000001), part of the European Union - NextGenerationEU funding.

\bibliography{library.bib}
\bibliographystyle{apalike}
\clearpage

\appendix
\section*{Appendix}

\subsection*{Proof of Proposition 1}
\begin{proof}
Let us consider a sample with $iid$ elements $\lbrace y_1, \dots, y_{n_{iid}} \rbrace$, drawn from a population via simple random sampling, where $y_i$ is the variable for the $i$-th unit with expected value $\mu$ and variance $\sigma^2$. Let us consider also $\lbrace g(y_1), \dots, g(y_{n_{iid}}) \rbrace$ with $g(y)$ a generic monotone transformation of the income variable, induced by $g(\cdot): \mathbb{R}^+ \rightarrow \mathbb{R}$, that changes for each measure, having expected value $\gamma$ and variance $\phi^2$.
 Considering that a generic inequality measure can be expressed as $\theta=f(\mu, \gamma)$ with  $f(\cdot)$ a twice-differentiable function, $\hat{\mu}=\sum_{i=1}^{n_{iid}} y_i/n_{iid}$ and $\hat{\gamma}=\sum_{i=1}^{n_{iid}} g(y_i)/n_{iid}$, we can easily obtain estimator moments as $\hat{\mu} \sim [\mu, \sigma^2/n_{iid}]$ and $\hat{\gamma} \sim [\gamma, \phi^2/n_{iid}]$. Consider moreover that
 \begin{align*}
Cov[\hat{\mu},\hat{\gamma}]=\EX[\hat{\mu} \hat{\gamma}]-\mu\gamma= \frac{1}{n_{iid}}  (\EX[y \cdot g(y)]-\mu\gamma)=\frac{Cov[y, g(y)]}{n_{iid}}.
\end{align*}  
Let us define the population value of a  generic inequality measure $\theta$ as $f(\mu, \gamma)$, with $f(\cdot)$ a generic twice-differentiable function.
 By expanding the inequality measure estimator $\hat{\theta}$ as $f(\hat{\mu},\hat{\gamma})$, via Taylor's expansion around the population values and considering its expected value: 
 \begin{align*}
\mathbb{E}[\hat{\theta}]&=\theta+\frac{1}{2}f_{\gamma, \gamma}(\gamma, \mu) \mathbb{V}[\hat{\gamma}]+f_{\gamma, \mu}(\gamma, \mu) Cov[\hat{\gamma}, \hat{\mu}]+ \frac{1}{2} f_{\mu, \mu}(\gamma, \mu)\mathbb{V}[\hat{\mu}]+O(n_{iid}^{-2})\nonumber\\
&=\theta+O(n_{iid}^{-1})+O(n_{iid}^{-1})+O(n_{iid}^{-1})+O(n_{iid}^{-2})\nonumber\\
&=\theta+O(n_{iid}^{-1}),
 \end{align*}
where $f_{\gamma}=\frac{\partial f(\gamma, \mu)}{\partial \gamma}$ and $f_{\gamma\mu}=\frac{\partial^2 f(\gamma, \mu)}{\partial \gamma \partial \mu}$.
\end{proof}

\subsection*{Reaching results of Table 1 in a different way}

By considering the linearization proposed by \citet{graf2011use} and extended by \citet{vallee2019linearisation}, we can easily reach the same results set out in Table 1 in a different way. Let us recall the notation as $\mathcal{U}$ denoting a finite population of $N (< \infty)$ elements. Let $y_i$ be the income of the $i$-th unit, where $y_i \in \mathbb{R}^+$, $\forall i = 1, \dots ,N$ and $\boldsymbol y=(y_1, \dots, y_N)$ its population vector. A sample $s$ of size $n$ is drawn with a complex sampling design with probability of selection $p(s)$ such that $p(s) \geq 0$ and $\sum_{s \subset \mathcal{U}} p(s)=1$. Let us define $\boldsymbol{\mathds{1}}=(\mathds{1}_1, \dots, \mathds{1}_N)$ the vector of sampling indicator of unit $i$, taking value 1 if unit $i$ is in the sample and 0 otherwise. The first order inclusion probability of unit $i$ is $\pi_i$, where $\pi_i=\mathbb{E}(\mathds{1}_i)$, i.e. the expectation with respect to the sampling design and its population vector $\boldsymbol \pi=(\pi_1, \dots, \pi_N)$. $\pi_{ij}$ denotes the second order inclusion probability for $i\neq j$.

Consider $\hat{\theta}=\hat{\theta}(\boldsymbol{\mathds{1}}, \boldsymbol y)$ an estimator of $\theta=\theta(\boldsymbol y)$ with $\hat{\theta}(\boldsymbol{\mathds{1}}, \boldsymbol y)$ twice differentiable with respect to $\boldsymbol{\mathds{1}}$. \citet{graf2011use} shows that an approximation for $\hat{\theta}$ is

 \begin{align*}
\hat{\theta}  \approx \theta + \sum_{i \in \mathcal{U}}(\mathds{1}_i-\pi_i) \frac{\partial \hat{\theta}}{\partial \mathds{1}_i} \Bigr|_{\boldsymbol{\mathds{1}}=\boldsymbol \pi}  + \frac{1}{2}  \sum_{i \in \mathcal{U}}  \sum_{j \in \mathcal{U}} (\mathds{1}_i-\pi_i)(\mathds{1}_j-\pi_j) \frac{\partial^2 \hat{\theta}}{\partial \mathds{1}_i\partial \mathds{1}_j} \Bigr|_{\boldsymbol{\mathds{1}}=\boldsymbol \pi}. 
 \end{align*}
We can derive the approximation of the bias as
 \begin{align}
\mathbb{E}[\hat{\theta}]  \approx \theta + 0  + \frac{1}{2}  \sum_{i \in \mathcal{U}}  \sum_{j \in \mathcal{U}} (\pi_{ij}-\pi_i \pi_j) \frac{\partial^2 \hat{\theta}}{\partial \mathds{1}_i\partial \mathds{1}_j} \Bigr|_{\boldsymbol{\mathds{1}}=\boldsymbol \pi}. 
\label{bias}
 \end{align}
In the following, we derive the bias for GE$(\alpha)$ with $\alpha \neq 0,1$, however such a result can be extended to all the other measures considered. 

Let us recall from the manuscript the survey estimator of GE($\alpha$) as a function of two Horvitz-Thompson type estimators which, under the assumption of $w_i=1/\pi_i$, can be rewritten as
 \begin{align*}
\hat{\mu}=\frac{1}{N} \sum_{i \in \mathcal{U}}\frac{\mathds{1}_i y_i}{\pi_i} 
\quad \text{and} \quad
\hat{\gamma}=\frac{1}{N} \sum_{i \in \mathcal{U}} \frac{\mathds{1}_i y_i^\alpha}{\pi_i},
 \end{align*}
with GE($\alpha$) estimator defined as 
 \begin{align*}
\hat{\theta}_{GE}(\alpha)= \frac{n (n-1)^{-1}}{\alpha(\alpha-1)}  \bigg( \frac{\hat{\gamma}}{\hat{\mu}^{\alpha}} -1\bigg).
 \end{align*}
    By applying \eqref{bias}, its bias may be expressed with an approximate result as 
     \begin{align}
\mathbb{E}[\hat{\theta}_{GE}(\alpha)] - \theta_{GE}(\alpha) \approx \frac{n (n-1)^{-1}}{2\mu^{\alpha+1}(\alpha-1)} \frac{1}{N^2}  \sum_{i \in \mathcal{U}}  \sum_{j \in \mathcal{U}} \bigg(\frac{\pi_{ij}}{\pi_i \pi_j}-1 \bigg) \bigg( y_i y_j \frac{ \gamma(\alpha+1)}{\mu} - y_i y_j^{\alpha} -  y_i^{\alpha} y_j \bigg).
\label{biasge}
 \end{align}
Considering that variance and covariance of the Horvitz-Thompson estimator are defined as 
    \begin{align}
\mathbb{V}[\hat{\mu}]=\frac{1}{N^2}  \sum_{i \in \mathcal{U}}  \sum_{j \in \mathcal{U}} y_i y_j \bigg(\frac{\pi_{ij}}{\pi_i \pi_j}-1 \bigg), \\
%\mathbb{V}(\hat{\gamma})=\frac{1}{N^2}  \sum_{i \in \mathcal{U}}  \sum_{j \in \mathcal{U}} y_i^\alpha y_j^\alpha \bigg(\frac{\pi_{ij}}{\pi_i \pi_j}-1 \bigg) \\
Cov[\hat{\mu}, \hat{\gamma}]=\frac{1}{N^2}  \sum_{i \in \mathcal{U}}  \sum_{j \in \mathcal{U}} y_i y_j^\alpha \bigg(\frac{\pi_{ij}}{\pi_i \pi_j}-1 \bigg),
 \end{align}
 as stated by \citet[pg. 30]{arnad2017}, \eqref{biasge}  can be easily rewritten as 
     \begin{align}
\mathbb{E}[\hat{\theta}_{GE}(\alpha)] - \theta_{GE}(\alpha) \approx \frac{n (n-1)^{-1}}{\mu^{\alpha+1}(\alpha-1)}  \bigg( \mathbb{V}[\hat{\mu}]\frac{ \gamma(\alpha+1)}{2\mu} - Cov[\hat{\mu}, \hat{\gamma}] \bigg).
 \end{align}
Such analytical result coincides with the corresponding formula in Table 1 of the manuscript.

\end{document}